# DATA AGGREGATION ROUTING PROTOCOLS IN WIRELESS SENSOR NETWORKS: A TAXONOMY

Saeid Pourroostaei Ardakani

Allameh Tabataba'i University, Tehran, Iran

## ABSTRACT

*Routing in Wireless Sensor Network (WSN) aims to interconnect sensor nodes via single or multi-hop paths. The routes are established to forward data packets from sensor nodes to the sink. Establishing a single path to report each data packet results in increasing energy consumption in WSN, hence, data aggregation routing is used to combine data packets and consequently reduce the number of transmissions. This reduces the routing overhead by eliminating redundant and meaningless data. There are two models for data aggregation routing in WSN: mobile agent and client/server. This paper describes data aggregation routing and classifies then the routing protocols according to the network architecture and routing models. The key issues of the data aggregation routing models (client/server and mobile agent) are highlighted and discussed.*

## 1. INTRODUCTION

A wireless sensor network (WSN) consists of small and tiny computing devices that are scattered to collect and report ambient data. The network nodes are usually static and communicate through provided wireless channels that are limited (in terms of communication range), unreliable and vulnerable to environmental noises, signal reflections, wireless interferences and/or physical obstructions [42]. The main objective of WSN establishment is to provide low-cost ambient data collection services. The nodes usually are small and cheap with limited energy, computation, communication and storage resources that are able to perform only a set of basic computation and communication tasks. They measure ambient data and transmit the result to the consumer access point (sink) as it has less resource limitation. WSN architecture is generally classified as either flat or hierarchical. The flat network is formed by the nodes which are usually randomly scattered in the area, whereas a hierarchical network is formed by clusters or the groups of nodes [49].

The key benefit of WSNs is that they can be implemented almost anywhere without the need for any specific communication infrastructure. It allows a WSN to be deployed as an alternative to non-existent infrastructure (for cost effectiveness) or if the existing infrastructure is not appropriate to use. Owing to this, WSN technology is used in diverse applications like education, ambient monitoring, transportation and health [42]. For example, in the case of education, this technology can be used to make a safe and easy-to-use laboratory in which the students experience scientific concepts in details.

WSNs are considered as an application of ad-hoc networks [38]. Similar to ad-hoc networks, there is no specific infrastructure for WSN and the network is deployed in a self-organising manner without any centralised control. However, there are five differences between WSN and ad-hoc networks [22]:

1. WSNs are densely deployed using a large number of sensor nodes, whereas ad-hoc networks usually consist of a fewer number of nodes with sufficient resources to compute and/or communicate.

2. Sensor node resources such as energy and communication/computation power are weaker than ad-hoc node. In other words, sensor nodes are tiny, weak and cheap (i.e TelosB node [44]), whereas ad-hoc nodes are usually more mature (i.e smart phones and/ PDAs) and have stronger resources. Sensor nodes usually have restricted processing modules (CPU), so they are not able to process complex jobs. They are not able to maintain large-scale data because they have limited storage capacities. In addition, they are not able to frequently communicate over long distance wireless links as they have limited radio range and power resources to broadcast wireless signals.

3. Message broadcasting is usually used in WSNs as maintaining a global addressing pattern such as IP (to support peer-to-peer communications) is very expensive in terms of network resource consumption. On the other hand, an ad-hoc network is able to support local communications between any pair of nodes using IP-based communication protocols.

4. In contrast to ad-hoc, WSN avoids collecting and transmitting redundant data as it increases network resource consumption.

5. WSN nature of dynamicity is different from ad-hoc network as sensor nodes are usually stationary in most applications.

WSN routing is in charge of interconnecting sensor nodes via either single or multi-hop links. It includes the procedure of route discovery, establishment and maintenance. The purpose of WSN routing is to forward data packets from event regions to the sink. However, routing raw data packets through wireless communication links from source regions to the sink increases network resource consumption and consequently reduces the network lifetime. This means that the sensor nodes would consume a great amount of network resources mainly-energy, if they need to forward each sensed data sample to the sink.

Data aggregation is a technique to combine data packets. This has the potential to eliminate meaningless/redundant data and reduce the number/size of transmissions. Hence, data aggregation technique can reduce network energy consumption if it is used in a WSN. This technique combines the data packets using an aggregation function (e.g Average, Maximum, Minimum, Count, Median, Rank, Standard Deviation, Variance and Sum) into a single one to transmit. It would result in reductions of transmissions and consequently decreasing the communication costs, bandwidth utilisation, network congestion, energy consumption and network delay in WSN routing.

WSN data aggregation routing makes communication paths between source nodes and the sink to aggregate and forward the network traffic. Resource conservation (mainly-energy), maximizing the number of captured data and minimizing data collection delay are three key issues that need to be considered in data aggregation routing [37]. Client/server and mobile/agent are two potential forms of data aggregation routing in WSN [47], [10]. Client/server lets the intermediate nodes to collect and aggregate data packets from the event region to the sink, whereas mobile agents are forwarded throughout the network to capture and collect data in mobile agent data aggregation routing model.

In the reminder of this article, Section 2 introduces WSN routing. It highlights the key issues which should be considered to design WSN routing protocols. Section 3 introduces data aggregation routing in WSN according to client/server and mobile agent models. The key design issues of each model are explained based on the protocol reactivity and/or network architecture (flat and hierarchical). Section 4 concludes data aggregation routing in WSNs and compares then client/server and mobile agent data aggregation routings.

## 2. ROUTING IN WSN

Routing in WSN utilizes a convergence pattern to forward data packets from source nodes to sink via either single or multi-hop links. Sensor nodes may need to forward network traffic through multi-hop links as they usually have limited communication abilities which do not allow direct communications. As a result, WSN routing needs to efficiently route the data packets from source regions to sink based on the network characteristics, node capacities and application requirements.

There are two components in WSN routing that work in parallel to route the network traffic: the protocol mechanism and the routing matrix [8]. The protocol mechanism focuses on data transmission scheme, data forwarding, routing information storage, packets characteristics and path discovery. The routing matrix works upon the protocol mechanism and its objective is path selection amongst the available ones. Route matrix returns the optimal paths if multiple paths are available. It is in charge of making the routing decisions according to the routing parameters such as consumed energy, path hop count, communication signal strength and loop avoidance [28].

The paths are established in WSNs in two schemes: Address-Centric (AC) and Data-Centric (DC). In the former, the nodes consider the address of next hop nodes to forward network traffic, whereas in the latter the routes are established using an attribute-based naming that specifies the properties of data over the wireless links. AC protocols do not offer benefits in WSNs because there is no global addressing scheme such as IP in WSN. In fact, lack of global addressing scheme in WSNs limits address-centric communications into local area in which sensor nodes are aware of each other ID address. Moreover, dense, dynamic and/or random WSN deployment complicates acquiring ID address of nodes on multi-hop communication links. By this, WSNs use DC routing protocols to forward data packets. In data-centric routing, a data packet is forwarded if it is desirable for the next node.

Network topology change is an issue that have high impact on WSN routing. A node or communication route fails when the energy level at the nodes falls below the required threshold for being alive or maintaining the communication links. In fact, WSN routing becomes challenging as the nodes need to consider energy consumption, (wireless) connectivity and coverage to forward network traffic.

### 2.1. WSN Routing Design Issues

A set of distinct factors needs to be considered when designing WSN routing protocols. They depend on routing scheme, network architecture and node characteristics [2], [48], [3]. The key ones are explained below:

- Network architecture: it has a significant impact on routing to discover and establish route that are used to forward packets. Data packets are routed to higher levels of hierarchy such as cluster-heads when WSN is hierarchical, whereas they are directly/indirectly forwarded to the sink in flat WSN.

- Node placement: node placement is rooted in the network applications and/or the consumer requirements and has the potential to influence routing connectivity and coverage. Sensor nodes can be placed in two schemes: deterministic and non-deterministic. In the former, the sensor nodes get manually placed and pre-determined routes are used to report the network traffic, whereas they are randomly scattered and the paths are dynamically formed in latter.

- Energy: this highly influences routing performance in WSN. As the required power for wireless communications is correlated to distance, forwarding network traffic through shortest paths is highly desirable to conserve energy.

- Data delivery model: routing is influenced by the data delivery models that are continuous, query-driven, event-driven or hybrid. For example, single path routing is not recommended in continuous data delivery (i.e habitat monitoring) as transmitting all the packets continuously through a single/same path can drain the energy of nodes being used (bottleneck). Owing to this, multi-path or hierarchical routing protocols are utilised to forward the network traffic through a set of variant paths or intermediate nodes that are able to eliminate redundant data.

- Node capabilities: nodes capability and functionality influences routing design and performance. For example, source nodes may forward data packets to the intermediate nodes such as cluster-heads which are able to perform in-network data aggregation.

- MAC protocol design: MAC protocols affect the routing performance as they are responsible for wireless link availability. For example, the link availability and consequently communication connectivity might be influenced by MAC protocols. Moreover, energy conservation can be enhanced in WSN if MAC protocols eliminate idle-listening energy consumption. They allow nodes to wake up when they need to send or receive network packets and then go to sleep if they have nothing to do.

- Data aggregation: as source nodes may forward a large amounts of redundant data measured from overlapped areas and/or similar events, data aggregation techniques are used by routing protocols to reduce the size and/or the number of (similar and/or redundant) data packets.

## 3. DATA AGGREGATION ROUTING IN WSN

Data aggregation routing aims to transmit a summarised scheme of sensed data (without losing data meaning and accuracy) in a convergent fashion to the consumer access point (sink). This leads to reduce transmission rate and consequently reduce network resource consumption. Data aggregation routing has two schemes [47], [10]: client/server and mobile agent. Client/server lets the intermediate nodes to collect and aggregate the forwarded data packets from the event region to the sink, whereas mobile agents are forwarded throughout the network to capture and collect data in the latter. In other words, the mobile agents need to migrate through the paths to capture and aggregate data samples at the source nodes and then return the results to sink.

## 3.1. Client/Server Data Aggregation Routing

Client/server data aggregation routing forms the paths in either flat or hierarchical. In the former, the nodes play same roles and the paths are established in a convergent manner from event regions to the sink. Apart from sink, intermediate nodes may perform in-network data aggregation if they receive multiple data packets. However, no node stays in charge of performing data aggregation process. On the other hand, the nodes may play different roles such as network bridge, intermediate aggregator or data consumer access point in hierarchical networks. The routes are usually established through intermediate nodes which perform data aggregation. Data packets are hierarchically aggregated and forwarded then from source nodes to the sink.

### 3.1.1. Flat Architecture

Flat data aggregation routing forwards data through minimum-cost paths which are formed from source nodes to the sink. The intermediate nodes might combine the received data packets when they are being transmitted to the sink. There are two schemes to form the routes in flat data aggregation: address-centric (AC) or data-centric (DC). The nodes consider the address of next hop nodes to forward network traffic in AC, whereas the routes are established using an attribute-based naming that specifies the properties of data over the wireless links in DC.

There are three classes to categorise flat client/server data aggregation routing [37]:

1. Push data aggregation routing: a subscription link initially is formed from the sink to source nodes to forward data. Source nodes, which receive the subscription links, become available to report data packets through the same links to the sink. The source nodes forward data packets until the subscription links are available. Push data aggregation routing constantly consumes network resources as source nodes would frequently transmit data packets as long as the subscription links are available. Moreover, the network resource consumption is increased as redundant or uninteresting data might be continuously reported to the sink. SPIN protocol (Sensor Protocols for Information via Negotiation) [24] uses push data aggregation routing. Under this protocol, meta-data is utilised instead of original data to establish the routes in flat. First, each source node introduces its own data packet to the single-hop neighbourhood by sending a message. A neighbour node replies back if it is interested in collecting. Then, the source node transmits a data packet to the neighbour node. The intermediate nodes collect and combine the received data packets and perform then a similar scheme to forward the aggregated data until the sink receives.

2. Two-phase pull aggregation routing: the source nodes heuristically establish shortest paths to transmit data packets to sink upon receiving the queries. It improves the quality and/or accuracy of data collection because data packets are forwarded according to the sink queries and not randomly or periodically. However, several round trip communications to establish the paths consumes network resources especially when network deployed is dense and sink queries are frequently changed. Direct diffusion [20] utilises a two-phase pull mechanism to collect, aggregate and report data. Under this protocol, data-naming technique is used to forward data packets. The objective of utilising the attribute-value is to reduce network resource consumption by eliminating unnecessary data processing and communication. This means that a data packet is forwarded if it matches the query attribute-values. Data processing and communication is reduced as only the nodes that have interesting data for the sink or are able to establish a link to the source regions perform routing. First, the sink sends out the queries consisting of data attribute values. These values such as data type and geographical area describe the data samples which the sink is interested in collecting. The query messages are occasionally broadcasted to refresh the route availability. The intermediate nodes update their routing tables upon receiving the query messages. The routing information is used to perform in-network data aggregation and form the return paths. Query message broadcasting is performed until nodes that have interesting data receive. The nodes may need to select the best available path because they receive a number of similar messages though variant routes. A set of parameters such as end-to-end delay or hop-count is considered to form the best path. The best path (called gradient) is used to forward data packet to sink. The nodes also may use other routes as alternatives if the current path (gradient) fails. The nodes living on the selected path combine and forward data packets until the sink is reached.

3. One-phase pull aggregation routing: a shortest path is formed between the sink and source nodes if the queries meet interesting data to report. This technique offers a high overhead data aggregation routing because the sink needs to collect location information of source nodes that have interesting data. An one-phase pull data aggregation routing protocol is proposed by [23] in which sink propagates interest packets to the network to establish shortest paths namely gradients to the source nodes. Each source node that receives query packets selects the minimum delay path (minimum hop count) to forward data packet if the query requirement is met.

Flat data aggregation routing needs to deal with the following drawbacks: (1) high overhead of establishing shortest paths especially in the case of large and dense networks, (2) simultaneous access to the wireless channels by the nodes to forward data packets results in increasing message failure and network congestion, (3) message re-transmission (due to collision and congestion) enhances energy consumption, (4) data aggregation accuracy and robustness is reduced due to data packets collision and congestion, (5) the routing paths may offer variant end-to-end delays which influence data freshness when data packets are forwarded through different routes (with variant hop count) [3].

### 3.1.2. Hierarchical Architecture

This model of data aggregation hierarchically forms an infrastructure to collect, combine and report data packets. Utilising an hierarchical infrastructure for data aggregation aims to resolve most flat data aggregation drawbacks. Data packets are not directly forwarded to the sink but they are routed to intermediate nodes which stay in charge of performing in-network aggregation. They get aggregated earlier in hierarchical networks as compared to flat. In fact, aggregator nodes hierarchically combine data packets in hierarchical routing instead of random aggregation which are performed by the nodes if reside on a joint path in flat routing. Hence, this reduces the number of relay nodes and consequently results in reduction of network traffic and congestion [19]. By this, message collision and end-to-end delay is reduced, however, data collection accuracy in increased. The benefits of hierarchical routing are outlined as below:

1. In-network data aggregation: this model of routing offers in-network data aggregation. Data packets are transmitted from source nodes to higher levels of hierarchy (i.e leader and/or cluster-heads) to aggregate. It leads to reduce the number of transmissions and network congestions. In other words, reducing the number of data packets results in reduction of message collision in hierarchical routing. Clustering, for example, is a technique which is commonly used to establish hierarchical infrastructures in WSN. In a clustered WSN, data packets usually are transmitted from source nodes (cluster members) to cluster-heads to collect, aggregate and/or transmit. Aggregated results are reported by cluster-heads to sink via either direct or indirect links.

2. Increasing the message delivery ratio: the probability of message failure/collision would be reduced in hierarchical routing as the network traffic decreases. In hierarchical routing, a set of specific nodes stays in charge of communication for a group of nodes. Indeed, a data sample is not directly forwarded to the sink by a source node. It reduces the number of nodes which try to access the wireless channels and the number of forwarding data packets. In consequence, message collision/failure is reduced, resulting in increasing the message delivery ratio.

3. Fair channel allocation: wireless communication channels can be efficiently managed in hierarchical network as contention-free scheduling is supported. Contention-free scheduling allocates wireless channel according to the nodes hierarchy level or location in advance of communications. It has the potential to increase the fairness of channel allocation and consequently decrease packet collisions as compared to contention-based scheduling which is increasingly used in flat networks.

4. Uniform energy consumption: in hierarchical routing, there is no centralised data processing but this is performed in a distributed manner. Hence, in-network tasks are hierarchically processed and analysed at the intermediate nodes. It leads to balance network loads resulting in increased network lifetime. On the other hand, a number of bottlenecks might arise in a flat network by the nodes which individually attempt to forward data packets. This results in non-uniform energy consumption that increases the risk of node failures in flat routing.

5. Routing delay reduction: communication delay is reduced in hierarchical routing because of utilising parallel links to report data samples. Besides, receive (queuing) and access delays [6] are reduced due to decreasing the number of messages and network traffic in hierarchical routing.

A hierarchical infrastructure can be formed in either static or dynamic fashion. The nodes are allocated by the network roles mainly intermediate aggregators in advance of network deployment if this is static, whereas they are reactively selected and the hierarchical infrastructure is dynamically formed in dynamic form. Dynamic hierarchical infrastructure is widely used because of frequent topology change and random deployment in WSNs.

Static hierarchical client/server data aggregation is easy and low-cost to set up as the intermediate aggregators are selected in prior to network deployment. The nodes do not need to consume network resources to dynamically form the infrastructure. Stronger nodes, in the terms of having sufficient computation, communication and storage resources, are positioned in efficient locations in which most possible number of source nodes are covered for data collection. It is commonly used in laboratory experimental situations because of their relative ease of installation. It allows source nodes to transmit data packets to the intermediate aggregators through energy efficient and low delay links. However, this scheme of hierarchical client/server data aggregation is not applicable for most applications of WSNs because of network topology change and random deployment.

Dynamic hierarchical data aggregation infrastructure is reactively formed in an ad-hoc manner by considering the available resources and capabilities of nodes such as remaining energy and/or coverage degree. The network is partitioned into a set of hierarchy levels (i.e clusters) that are managed by either a single or multiple nodes (i.e leaders or cluster heads). The nodes can be selected using voting and/or probability algorithms [4]. The key duty of these nodes is to collect, combine and forward data samples. Node density, distribution pattern, connectivity and coverage degrees are the issues that need to be considered to establish dynamic hierarchical infrastructure.

There are a set of different techniques that are used to form hierarchical infrastructures in WSNs. Aggregation tree, clusters and chain are the most commonly used ones that are explained as below:

- Aggregation tree: this is established wherein a packet is hierarchically reported from each node to its parent for in-network data aggregation. The objective is to minimise resource consumption and maximise data collection rate [5]. The tree is formed using source nodes which report interesting data to the sink. TAG (a Tiny AGgregation service for ad-hoc sensor networks) [30] forms a tree infrastructure to collect and combine ambient data. At the beginning, the sink sends a level discovery message to assign a level label to network nodes. Each node increases its own level value by one and then forwards the message to the next hop if the message is received. This is continuously performed until all nodes get a level value. Then, the nodes forward data packets if they detect an available path to their upper level nodes. This procedure is repeated until the sink captures the aggregated result. TINA (a scheme for Temporal coherency-aware In-Network Aggregation) [41] utilises a similar mechanism to establish a tree infrastructure wherein data aggregation is performed. TiNA differs from TAG as it utilises temporal coherency tolerance. Under TiNA, the source nodes transmit data values if they differ from the pervious data reports. By this, "tct" parameter is added to the queries to show the consumer preference tolerance degree. By this, a data sample is forwarded if differs with the last value greater than "tct". As the source nodes do not transmit all the measured data, TiNA reduces the network energy consumption. This is also supported by the empirical results presented in [41].

- Clustering: the network is partitioned into a set of groups named clusters using clustering technique. There are two schemes to form the clusters: address-centric and data centric. By this, the nodes that are similar in location or communication characteristics are grouped as a cluster. The nodes that reside in a cluster are named Cluster Members (CMs). Among all CMs, a single or multiple nodes stay in duty of managing the cluster. They are called Cluster Heads (CHs). CHs usually collect and combine intra-cluster data samples. Low-Energy Adaptive Clustering Hierarchy (LEACH) [18] is an address-centric routing algorithm that supports data aggregation. LEACH has two phases: setup and steady-state. The setup phase forms the clusters, whereas the steady-state forwards network traffic to sink. LEACH utilises a distributed random algorithm to select CHs. This is periodically performed and leads each CH stays in the duty for a particular round based on a value (P). This means that there is no chance for a CH to get the same role up to P next rounds. TDMA (Time Division Multiple Access) [31] is utilised by source nodes to collect and report the network traffic and avoid intra-cluster collision. In addition, CDMA (Code Division Multiple Access) [7] is used by CHs to forward the aggregated result to the sink and avoid inter-cluster interference. Forwarding the network traffic in unicast (instead of multicast) reduces energy consumption in LEACH. CLUstered Diffusion with Dynamic data Aggregation protocol (CLUDDA) [37] diffuses the sink queries into a clustered network in which the CHs are in charge of performing in-network data aggregation. The queries include data collection information such as data type and aggregation function. Each CH that meets the requirements collects and aggregates intra-cluster data samples and then forwards the result to the sink. CLUDDA is a data-centric protocol and allows the data consumer to partially collect and aggregate data samples from each region of network in which data is desirable. It reduces energy consumption as data aggregation is performed by only a selective set of CHs (instead of all CHs) that match the interest packet requirements.

- Aggregation chain: this forms a hierarchical infrastructure for data aggregation wherein source nodes report data samples to the sink. The chain is managed by a leader node which has the responsibility of collecting data samples from source nodes and reporting the aggregated result to the sink. By this, data samples are hierarchically aggregated and forwarded then from the source nodes to the leader. The aggregated result is forwarded by leader(s) to the sink via direct or indirect links. PEGASIS (Power-Efficient GAthering in Sensor Information Systems) [27] establishes a chain infrastructure for data aggregation. This protocol hierarchically forms a chain-based infrastructure to route data packets. The chain leader nodes are selected by considering a set of distinctive parameters such as residual energy level or location information. Then, source nodes utilise a greedy algorithm to forward data samples to next hop based on the distance to leader(s). This means that they forward a data packet if the next hop node is closer to the leader. Data packets get aggregated at each node and forwarded then to the next one until the leader is reached. Finally, the leader reports the aggregated result to the sink. However, a new leader is selected and a new chain is formed if the leader that is on duty fails. This protocol differs from LEACH as avoids periodical infrastructure reforming (i.e re-clustering). Due to this, and according to the simulation results in [27], PEGASIS outperforms LEACH in terms of network lifetime. However, the cost of PEGASIS is increased over long period. This means that network resource consumption is increased in PEGASIS as sensor nodes need to collect information which is required to re-form the chain and re-select the leader. In addition, frequent data report by the leader makes them bottleneck. Besides, multi-hop transmissions from source nodes to the sink increase end-to-end delay in PEGASIS. To resolve this, Hierarchical-PEGASIS [27] is proposed. This aims to reduce delay as source nodes forward network traffic using parallel links to the sink. For this, [27] utilises two techniques: signal coding (e.g CDMA) and transmitting spatial separated data. In the former, a tree infrastructure is formed that gets rooted in the sink. This allows the nodes to forward data packets in parallel. However, CDMA is utilised by each parallel transmission to avoid collision. The latter allows the nodes report data to the leader(s) if they are in the same region. The source nodes are spatially clustered and transmit then their data samples to the leader(s) in parallel.

Hierarchical client/server data aggregation needs to deal with the following issues [29]: (1) the overhead of hierarchical infrastructure establishment/maintenance: this enhances the network resource consumption (mainly-energy) that consequently results in the reduction of network lifetime. Sensor nodes need to consume a great deal of energy to establish or re-establish the hierarchical structure when network topology or density changes, (2) Leader/CH bottlenecks: the intermediate aggregators on the hierarchical infrastructure such as CHs stay in the duty of managing in-network jobs including computation and communication tasks more than other nodes. Hence, the aggregator nodes have a higher chance to fail (due to running out of energy) as they need to manage a great number of communication and computation tasks. Table 1 highlights and compares the key features of flat and hierarchical data aggregation routings in WSN.

Table 1. Flat vs. Hierarchical

| Parameters | Flat | Hierarchical |
|---|---|---|
| Aggregators | Any node | Intermediate aggregators |
| Node fail | Network partition | Local disconnect |
| Congestion | High | Low |
| Collisions | High | Low |
| Cost | Low | High |
| Node heterogeneity | No impact | Makes the nodes role |
| Delay | High | Low |

## 3.2. Mobile Agent Data Aggregation Routing

This model of data aggregation routing utilises Mobile Agent (MA) technique to collect and aggregate data from source nodes. The key objective is to increase data aggregation accuracy and performance and reduce network resource consumption. This section briefly describes mobile agents structures, capabilities and benefits. A set of mobile agent routing protocols is provided to highlight routing issues and techniques that need to be considered in WSN data aggregation.

### 3.2.1. Mobile Agents: A Brief Review

A Mobile Agent (MA) is a piece of software that has mobility to autonomously perform distributed computing tasks [12], [16]. There are two techniques to provide MA mobility: mobile code and remote objects [9]. In the former, code migration is provided and managed by a software framework (i.e Telescript Development Environment) [33], whereas the latter (i.e Aglets) focuses on Remote Object Invocation (ROI) that allows remote access to the information/object with respect to the system transparency [25].

Programmability is the key feature of mobile agent as compared to regular computer software. It provides the ability of collecting and processing information, and then performing the best-fitted services for the consumer. Programmability enhances the performance of computing systems in which the MAs are used to intelligently manage resources [26]. Based on the definition domain, the MAs support programming in application, middleware or network layers [1]. In the application layer, MAs are usually used to enhance the flexibility/efficiency of application design by propagating the consumer requirements in an autonomous manner. In the middleware layer, the MAs can be used to enhance dynamicity of network services such as data aggregation and/or query-based information retrieval. They also are used to improve the intelligence of network layer services such as smart multi-hop routing.

### 3.2.2. Mobile Agents Structure and Benefits for Data Aggregation

The MAs can be programmed to perform data aggregation in WSNs. They move throughout the network to capture and aggregate data samples which need to be reported to the sink. A mobile agent usually consists of four elements: identification, itinerary, data space and method [34]. The identification maintains the general information of MA such as serial number and/or dispatcher's ID. Itinerary provides the migration information such as current location, traversed paths and/or destination address. Data space keeps the required and/or collected data (i.e aggregated result) during the MA migrations. Method provides the required code/function (i.e aggregation function) that is used by MAs during the migrations between the computing devices. As a result, the MAs would be able to visit the source nodes one by one using the itinerary information that can be provided proactively or reactively. They aggregate captured data at each node using the aggregation function. Aggregated results are maintained and/or updated at data space of MAs until they are delivered to the sink.

According to [32] and [34], the MA model of data aggregation offers nine advantages in comparison to client/server as below:

1. Utilising the MA technique decreases transmission rate in WSN. MA routing forwards the executable sink queries (MAs) to the source nodes to collect and combine data samples, whereas client/server routes a large amounts of raw data from the source nodes to the aggregators/sink. For example, let we assume that a set of particular photos taken by wireless camera sensors need to be collected. In client/server, camera sensors report all their photos to either sink or aggregators for aggregation and/or processing/analysis. On the other hand, a MA can be programmed to move throughout the network to collect the photos which are interesting and met the user requirements. Hence, the number of transmissions is reduced in MA data aggregation.

2. Reducing the number of transmissions results in decreasing network resource consumption mainly energy.

3. The network traffic and transmission rate is reduced if MA data aggregation is used. This results in reduction of collisions.

4. Reducing network traffic leads to decrease of data aggregation delay in WSNs.

5. MA data aggregation performance is not dependant to the network size. This increases the scalability of data aggregation.

6. MA improves the extensibility of system as it has the potential to carry task-adaptive processes which extend the capabilities of the system.

7. MA has the ability to improve the stability of network as it is able to support offline (asynchronous) message delivery.

8. Bandwidth utilisation is reduced in MA model of data aggregation because of transmitting Mas instead of data packets throughout the network.

9. Load balancing is enhanced in MA data aggregation due to uniform network load distribution.

MA migration itinerary planning is a challenging issue in MA data aggregation. It is clearly related to the Travelling Salesman Problem (TSP) in which optimal itineraries are established for salesmen to follow. Although solving TSP (and similarly MA itinerary planning) is practical when the number of nodes to visit (i.e cities) is small, the problem is in general NP-complete. However, there are three key differences between TSP and MA itinerary planning: (1) TSP needs to visits all the nodes (i.e cities), whereas MA itinerary planning only visits the nodes which are desirable for the consumer. (2) there is a single salesman which travels through in traditional TSP, whereas MA itinerary planning focuses on routing multiple MAs throughout the network. (3) TSP assumes global knowledge, whereas MA itinerary planning in WSN does not.

MA migration itinerary planning needs to consider three issues to discover/establish optimal routes [32]: (1) Minimising journey delay: this leads to enhance data freshness as data samples are collected with minimum delay. (2) Reducing network resource consumption (mainly energy): MAs need to move through short, low cost and efficient energy use routes as WSNs are highly resource constraints. (3) Maximising data sample rate: data collection robustness is increased if greater number of data samples is captured.

MA itineraries are usually designed in two fundamental schemes depending on where the routing decisions are made: proactive and reactive [45]. MAs utilise the itinerary that is created by sink or data consumer in advance of migration in the former, whereas MAs are routed on-the-fly according to the acquired routing information that are dynamically collected in the latter. However, the MAs may use hybrid routing in which proactive and reactive itinerary planning are combined.

### 3.2.3. Proactive MA Data Aggregation Routing

A pre-defined itinerary is set for MA migration in proactive routing. This reduces the cost of routing, network resource consumption and delay [13]. As the migration paths are defined (by sink and/or data consumer) priori to migration, the overhead of route discovery and establishment can be reduced. This means that routing overhead is reduced as on-the-fly route computing in not required. In addition, the pre-defined itineraries return the shortest paths that reduce MA migration delay. Proactive MA routing can be used in laboratory experimental situations because of its relative ease of installation. However, it is not applicable on most WSN data aggregation applications as the MA dispatcher (sink) should be aware of source nodes address, location, source events and the possible shortest paths between them. Sensor nodes may be randomly distributed in harsh and/or out of reach area like the ocean. Hence, collecting required routing information to design proactive itineraries would be difficult, expensive and/or impossible.

*Mobile Agent Based Wireless Sensor Network (MAWSN)* [14] proactively allocates a migration itinerary to a single MA to follow. The MA utilises a pre-defined routing which consists of three parts: (1) the first node id: this shows the beginning of the journey, (2) intermediate nodes id: the list of nodes which should be visited during the journey, (3) the last node id: this returns the end of migration. According to the itinerary, the MA is moved from the sink to the first node through a proactive path to start the journey. Then, the MA selects the source nodes one by one from the itinerary list to visit. Distance to the sink ranks the nodes in advance of the MA trip. The MA moves to a node if it is closer to the sink. The procedure is performed until the MA visits all the source nodes in the list. At the end, the MA moves to the last node to return the results through a reserved path to the sink.

*Mobile Agent Distributed Sensor Network (MADSN)* [36] moves a MA using a proactive routing map to collect and aggregate data samples similar to MASWN. The difference of MAWSN and MADSN is that, the latter utilises MRI (Multi Resolution data Integration) [34] technique to reduce data redundancy. Using this, the nodes avoid reporting data samples if they are redundant. In other words, MADSN avoids visiting the source nodes that have redundant data. This reduces the cost of data collection comparing to MAWSN. Hence, it is clear that MADSN offers benefits if it is used in clustered network because CHs discard redundant data using MRI.

*Mobile Agent Directed Diffusion (MADD)* [13] moves a single MA throughput the network in a hybrid manner for data aggregation. The MA uses itinerary similar to MASWN. The MA reactively migrates between first and last nodes that are proactively selected by the sink. It means that, the MA is moved to the first node by sink and then dynamically selects the closest node until it is reached by the last node. At the end, the MA returns to the sink from the last node through a reserved path. There are differences between MADD and MASWN: (1) MADD reactively visits the nodes between the first and last visiting node. It dynamically selects the closer source node to its current location to move in each migration. (2) MADD selects the farthest source node from sink to start the MA journey. The reason is that the author believes moving empty/lightweight MAs through longer routes in early migrations reduces network resource consumption. In fact, starting the migration from farther nodes using a MA in which data part is empty reduces communication overhead (size × communication distance) that has a high impact on energy consumption. (3) MADD discards the method part when the last source node is visited. This results in reduction of the network traffic.

### 3.2.3. Reactive MA Data Aggregation Routing

Reactive MA data aggregation derives MAs to visit the source nodes through paths that are dynamically formed in flat or hierarchical [45]. The routing information at each node is used to establish the routes. For example, MAs select the closest node to migrate in the next using Received Signal Strength Indicator (RSSI) [46] technique which estimates the distance between the wireless nodes. Reactive MA routing is not influenced by network topology change. However, on-the-fly information collection and reactive route establishment increases data migration delay.

*GCF (Global Centre First) and LCF (Local Closest First)* [35] move a single MA in flat into the event region(s) for data aggregation. A single MA moves to visit the source nodes (via the shortest path) if they are close to the centre of data region in GCF, whereas LCF utilises a routing algorithm in which the MA is moved to the next source node if it is the closest one. The complexity of these two algorithms is comparatively low and they are easy to implement. However, data aggregation cost and delay depends on the network size and this is increased if the network becomes large and dense [32]. In addition, the performance of these protocols highly is influenced by the current location of MA. For example, the sink should know the centre of data region to move the MA if GCF is used. Although it is not critical in centralised event distribution model, the MA migration cost is highly increased when random event distribution model is used especially in the network deployed is large and dense.

*IEMF (Itinerary Energy Minimum for First-source-selection) and IEMA (Itinerary Energy Minimum Algorithm)* [15] move MAs via minimum cost routes to aggregate and report data. The key objective is to decrease MA migration overhead in IEMF. This selects minimum consumed energy paths to route the MA. IEMF allocates an estimated cost value to each route that is established to an event region. According to the cost value, it selects the closest node that resides on minimum cost link to migrate. LCF differs from IEMF as this selects the closest node for migration, whereas IEMF utilises the estimated cost value to select the link. Utilising an iterative process in IEMF to select the next hop nodes forms IEMA. First, each available tie to data region is assigned by a cost value. Then, the value is iteratively updated if the real cost is measured. Indeed, IEMA considers a number of available links to event regions in an iterative manner to find out the route in which MA migration cost is minimised.

*The Near-Optimal Itinerary Design algorithm (NOID)* [17] utilises multiple MAs which independently travel throughout flat networks to collect and aggregate data samples. This enhances the parallelism of data aggregation routing that consequently reduces delay. This means that this protocol reduces data aggregation delivery time as a number of MAs in parallel collect and report data. The migrations are started from the sink to data regions via the established paths. Each route is extended in a greedy manner to minimise a cost function in which hop count and node energy level is considered. NOID allocates a cost value to each link. It allows the MAs to select the closest node residing on the minimum cost link to move. In other words, the MAs move through links which minimise journey hop count (minimum delay) and have sufficient energy to guide them to source regions. NOID also considers the amount of collected data at each nodes to control MAs size. Forwarding MAs without considering the MA size increases the network traffic and network resource consumption. For this reason, NOID monitors the data part of MAs at each node and avoids continuous node visit. In fact, an MA stops to visit nodes and returns to the sink if it becomes heavy. However, MA migrations to overleaped areas and capturing redundant data samples are the drawbacks of NOID. The MAs only consider the address of nodes instead of their available data during data aggregation journeys. In consequence, they visit overleaped area and capture redundant data. In addition, the migration cost of multiple MAs is increased if the number of data regions rises.

*Genetic Algorithm based Multiple mobile agents Itinerary Planning (GA-MIP)* [11] utilises Genetic Algorithm (GA) to compute routes for multiple MAs. To avoid collision and overlapped migration, GA-MIP shares a random itinerary as the initial gene between all MAs. The route map has two vectors: sequence and group. The former keeps the list of source nodes that need to be visited, whereas the latter indicates the number of source nodes that should be visited by each MA. This means that each element of group vector specifies the number of nodes that are listed in sequence vector and need to be visited by a MA. The vectors are updated/trained using the GA algorithm. GA operators such as crossover and mutations increase the variety of other route maps (gens). These are evolutionary created based on the network topology change. At each iteration, the best-fitted route maps are selected using the selection operator. This procedure is repeated until the most efficient itinerary is achieved for the MAs to follow. The drawback of GA-MIP is delay. This is increased because initial gene optimisation during data aggregation procedure is required. For this reason, GA-MIP is not suitable for real-time or time-sensitive data aggregation applications.

*Tree-Based Itinerary Design (TBID)* [21] utilises a number of spanning trees (SPTs) to combine and report data samples over a zoned network. Under this protocol, the trees are established from the single-hop neighbourhood of sink. Then, an MA for data aggregation is assigned to each tree. After this, a set of concentric zones is formed around the sink. Radius of each zone is N ×(maximum radio range of node)/2 in which N is the zone number. Each node residing in the first zone (single-hop neighbours of sink) starts to establish a spanning tree with the source nodes. To form a tree, the nodes use a greedy algorithm. Using this algorithm, each source node interconnects to the next node which resides in the outer zone. This procedure is repeated until source nodes in the last zone (most outer one) are reached. Finally, the MA data aggregation trip is started to collect data samples residing on the tree. Each MA visits all the nodes in each zone and move then to the outer zone to continue the procedure. The same path is used by the MAs to return to the sink. The key drawback of this protocol is that the trees are proactively established. This enhances energy consumption if the network topology frequently changes. In addition, data aggregation cost is increased in TBID if the network is deployed dense and large [39] proposes a reactive mobile agent data aggregation protocol that is particularly designed for object tracking. It uses two types of MAs: master and slave. A Master MA is responsible for detecting the event regions and reporting aggregated results to the sink, whereas a slave MA collects data samples within a region to report to its master. According to it, a master MA is forwarded from the sink to the closest source node (minimum hop count) of each event region. Each master MA detects the current status of the event (i.e object location and direction) using the collected information at the source node. Then, the slave MAs are routed to collect and combine data samples in the data region. In object tracking application, for example, slave MAs move to the sensor nodes that are able to dominate the movement space of the object. They report collected data samples to master MA in two schemes: Threshold-Based (TB) and Distance-Based (DB). TB allows slave MAs report if the object meets the required speed or direction, whereas DB lets slave MAs return data if the object move beyond the allowed distance. Slave MAs collect and aggregate data samples and then report the results to the master MA. Master MA collects and aggregates the information from its salve MAs and then return to the sink to report. The drawback of this algorithm is the performance cost. The routing performance cost increases when the event/objects are distributed according to random event distribution model. For this, a greater number of MAs is required to report data from the event sources. It increases network resources consumption. Besides, the accuracy of data aggregation is decreased as the MAs may move into overlapped area and collect redundant data. Agent-based Event driven Route Discovery Protocol (AERDP) [40] (and [43]) aims to resolve the drawbacks by establishing a non-overlapped hierarchical infrastructure (clusters) for MAs to migrate. The clusters are interconnected through a spanning tree which is rooted in the sink. Similar to [39], a master MA stays in the duty of collecting data from slave MAs at each region. The overhead/complexity of this algorithm highly depends on the size and distribution model of event regions. It increases when the event regions are broadly formed with numerous and/or random events.

Table 2. Client/server vs. Mobile agent data aggregation routing

| Protocol | Network Architecture | Mechanism | Key Advantage | Key Drawback | Number of MAs |
|---|---|---|---|---|---|
| LCF | Flat | Closest node | 1- Ease of implementation 2- low overhead | 1- AC node visit 2- Increased delay 3- Location dependant | Single |
| GSF | Flat | Centre of event regions | 1- Ease of implementation 2- low overhead | 1- Increased delay 2- Location dependent | Single |
| NOID | Flat | Closest node with enough energy | Reliable migrations via energy-aware links | Overlapped data collection | Multiple |
| IEMF & IEMA | Flat | Minimum cost links (energy and distance) | Reducing migration costs | Increased delay | Single |
| GA-MIP | Flat | GA for optimising links | Best-fitted routes according to the requirements | Increased delay of route planning | Multiple |
| TBID | Hierarchical | Tree infrastructure | Reducing MA random walks | 1- High cost of updating tree infrastructure 2-Great number of trees for large network | Multiple |
| [39] | Hierarchical | Master/Slave MAs | Reducing MA migration length | 1- Overlapped data collection 2-High cost for random distributed event sources | Multiple |
| AERDP | Hierarchical | Master/Slave MAs | Non-overlapped migrations | High cost for random distributed event sources | Multiple |

The introduced mobile agent routing protocols are highlighted and compared in Table 2.

## 4. CONCLUSION

Routing is used in WSN to route data samples from data regions to consumer access point (sink) based on distinctive parameters such as network architecture and application. A number of routing protocols are proposed in this field to improve routing stability, scalability and extendibility. They are classified according to routing features, techniques and objectives.

International Journal of Computer Networks & Communications (IJCNC) Vol.9, No.2, March 2017

WSN Routing is a challenging issues as it can be frequently influenced by network dynamism and/or topology changes. Residual energy level is the key parameter that usually lead to the network topology changes in WSNs. This affects node availability and wireless communication quality. Hence, it influences wireless communications and consequently routing performance. By this, the following objectives need to be considered by WSN routing:

1. The route stability needs to be enhanced by establishing/re-establishing communication links as quick as possible before any further topology change.

2. The reliability of routes needs to be enhanced by forwarding network packets via the nodes which have sufficient energy to communicate.

3. Network lifetime needs to be maximised by minimising energy consumption. It can reduce node/link failures that are caused by running out of energy.

4. A distributed scheme is required to process routing overhead. This avoids arising bottlenecks and consequently results in reduction of partial node failure probability

Data aggregation offers a set of benefits such as reducing network congestion and energy consumption in WSN routing by reducing size/number of transmissions. Data aggregation routing focuses on two schemes: client/server and mobile agent. In the former, the routes are formed between source nodes and the sink according to the network architecture. The paths guide data packets from the source regions to the sink through a set of intermediate nodes which perform in-network data aggregation. In MA routing, MA(s) move throughout the network via proactive/reactive paths to collect and report data. It offers a set of benefits specifically reducing network traffic, enhancing adaptability and autonomous computation as compared to client/server model. However, itinerary planning to establish efficient and low-cost paths for MAs is a challenging issue in MA data aggregation routing. Table 3 summarises and compares the key features of data aggregation routing in both schemes.

Table 3. Client/server vs. Mobile agent data aggregation routing

| Parameters | Client/Server | Mobile Agent |
|---|---|---|
| Communication | Uni/multi/broadcast (depending on architecture) | (parallel)Unicast |
| Parallel processing | Yes | Yes |
| Automaticity | No | Yes |
| Message Structure | Simple | Complex |
| Communication Overhead | Number of relay nodes (depending on architecture) | Number/length of MAs |
| Message Failures | Depending on architecture, traffic and/or energy | Depending on residual energy of nodes |
| Accuracy of data collection | Depending on data msg. failures | Depending on itinerary planning |
| Delay of data collection | Depending on traffic and/or architecture | Depending on itinerary and/or number of MAs |
| Key advantage | Simplicity in deployment | 1-Reducing network traffic 2-Adaptability |
| Key drawback | Message collisions/lost | MA complexity/security |

104

International Journal of Computer Networks & Communications (IJCNC) Vol.9, No.2, March 2017

## REFERENCES

[1]  Francesco Aiello, Giancarlo Fortino, Antonio Guerrieri, and Raffaele Gravina. Maps: A mobile agent platform for WSNs based on java sun spots. A workshop of the 8th International Joint Conference on Autonomous Agents and Multiagent Systems (AAMAS-09), Budapest, Hungary, 12 May, pages 25–32, 2009.

[2]  Kemal Akkaya and Mohamed F. Younis. A survey on routing protocols for wireless sensor networks. Ad Hoc Networks, 3(3):325–349, 2005.

[3]  Jamal N. Al-Karaki and Ahmed E. Kamal. Routing techniques in wireless sensor networks: A survey. IEEE Wireless Communications, 11(6):6–28, 2004.

[4]  Hitha Alex, Mohan Kumar, and Behrooz Shirazi. Midfusion: An adaptive middleware for information fusion in sensor network applications. Information Fusion, 9(3):332–343, 2008.

[5]  Mohammad Hossein Anisi, Abdul Hanan Abdullah, and Shukor Abd Razak. Efficient data aggregation in wireless sensor networks. International Conference on Future Information Technology (ICFIT'11), Singapore, Singapore, 13:305–310, September 16-18, 2011.

[6]  Saeid Pourroostaei Ardakani, Julian Padget, and Marina De Vos. HRTS: A hierarchical reactive time synchronization protocol for wireless sensor networks. Ad Hoc Networks, 129:47–62, 2014.

[7]  B. Zurita Ares, C. Fischione, and K. H. Johansson. Wireless Sensor Networks, volume 4373 of Lecture Notes in Computer Science, chapter Energy consumption of minimum energy coding in CDMA wireless sensor networks, pages 212–227. Springer Berlin/Heidelberg, Delft, The Netherlands, 2007.

[8]  Shadi Saleh Ali Basurra. Collision Guided Routing for Ad-Hoc Mobile Wireless Networks. PhD thesis, Department of Computer Science, University of Bath, October 2012.

[9]  Andrzej Bieszczad, Bernard Pagurek, and Tony White. Mobile agents for network management. IEEE Communications Surveys, 1(1):2–9, 1998.

[10] Pratik K. Biswas, Hairong Qi, and Yingyue Xu. Mobile-agent-based collaborative sensor fusion. Information Fusion, 9(3):399–411, 2008.

[11] Wei Cai, Min Chen, Takahiro Hara, and Lei Shu. Gamip: Genetic algorithm based multiple mobile agents itinerary planning in wireless sensor networks. The 5th Annual International Wireless Internet Conference, Singapore, March 1-3, pages 1–8, 2010.

[12] Yongtao Cao, Chen He, and Lingge Jiang. Energy-efficient routing for mobile agents in wireless sensor networks. Frontiers of Electrical and Electronic Engineering in China, 2(2):161–166, 2007.

[13] Min Chen, Taekyoung Kwon, Yong Yuan, Yanghee Choi, and Victor C. M. Leung. Mobile agent-based directed diffusion in wireless sensor networks. EURASIP Journal on Advances in Signal Processing, 2007:1–13, 2007.

[14] Min Chen, Taekyoung Kwon, Yong Yuan, and Victor C.M. Leung. Mobile agent based wireless sensor networks. Journal of computers, 1(1):14–21, April 2006.

[15] Min Chen, Victor Leung, Shiwen Mao, Taekyoung Kwon, and Ming L. Energy-efficient itinerary planning for mobile agents in wireless sensor networks. IEEE International Conference on Communications (ICC), Dresden, Germany, June 14-18, pages 1–5, 2009.

[16] Edison Pignaton de Freitas. Cooperative Context-Aware Setup and Performance of Surveillance Missions Using Static and Mobile Wireless Sensor Networks. PhD thesis, Halmstad University, 2011.

[17] Damianos Gavalas, Aristides Mpitziopoulos, Grammati Pantziou, and Charalampos Konstantopoulos. An approach for near-optimal distributed data fusion in wireless sensor networks. Wireless Networks, 16:1407–1425, 2010.

[18] Wendi Rabiner Heinzelman, Anantha Chandrakasan, and Hari Balakrishnan. Energy-efficient communication protocol for wireless micro-sensor networks. The 33rd Hawaii International Conference on System Sciences (HICSS'00), the Island of Maui, 4-7 January, pages 3005–3014, 2000.

[19] Fei Hu, Xiaojun Cao, and Carter May. Optimized scheduling for data aggregation in wireless sensor networks. International Symposium on Information Technology: Coding and Computing (ITCC 2005), Las Vegas, Nevada, USA, April 4-6, 2:557–561, 2005.

[20] Chalermek Intanagonwiwat, Ramesh Govindan, and Deborah Estrin. Directed diffusion: A scalable and robust communication paradigm for sensor networks. The 6th Annual International Conference on Mobile Computing and Networking (MobiCom '00), Boston, Massachusetts, August 6-11, pages 56–67, 2000.

[21] Charalampos Konstantopoulos, Aristides Mpitziopoulos, Damianos Gavalas, and Grammati Pantziou. Effective determination of mobile agent itineraries for data aggregation on sensor networks. IEEE TRANSACTIONS ON KNOWLEDGE AND DATA ENGINEERING, 22(12):1679–1693, 2010.
[22] Bhaskar Krishnamachari, Deborah Estrin, and Stephen Wicker. Modelling data-centric routing in wireless sensor networks. The 21st Annual Joint Conference of the IEEE Computer and Communications Societies (INFOCOM), New York, USA, June 23-27, 2(4):1–11, 2002.
[23] Bhaskar Krishnamachari and John Heidemann. Application specific modeling of information routing in wireless sensor networks. In the proceeding of IEEE international performance, computing and communications conference,, 23:717–722, 2004.
[24] Joanna Kulik, Wendi Rabiner, and Hari Balakrishnan. Adaptive protocols for information dissemination in wireless sensor networks. The 5th Annual ACM/IEEE International Conference on Mobile Computing and Networking (MobiCom' 99), Seattle, Washington, August 15-20, pages 174–185, 1999.
[25] Danny B. Lange, Mitsuru Oshima, Gnter Karjoth, and Kazuya Kosaka. Aglets: Programming mobile agents in java. Worldwide Computing and Its Applications (WWCA97), Lecture Notes in Computer Science, 1274, 1997.
[26] Sergio Gonza LezValenzuela, Min Chen, and Victor C.M Leung. Applications of mobile agent in wireless networks and mobile computing. Advances in Computers, 82:113–163, 2011.
[27] Stephanie Lindsey, Cauligi S. Raghavendra, and Krishna M. Sivalingam. Data gathering algorithms in sensor networks using energy metrics. IEEE Transactions on Parallel Distributed System, 13(9):924–935, 2002.
[28] Changling Liu and Jorg Kaiser. A survey of mobile ad hoc network routing protocols. Technical report, University of Ulm, 2005.
[29] Kai-Wei Fan Sha Liu and Prasun Sinha. Structure-free data aggregation in sensor networks. IEEE TRANSACTIONS ON MOBILE COMPUTING, 6(8):929–942, August 2007.
[30] Samuel Madden, Michael J. Franklin, Joseph M. Hellerstein, and Wei Hong. Tag: A tiny aggregation service for ad-hoc sensor networks. Fifth Symposium on Operating Systems Design and implementation (OSDI 02), Boston, MA, USA, December 9 - 11, pages 131–146, 2002.
[31] Jianlin Mao, Zhiming Wua, and Xing Wuc. A TDMA scheduling scheme for many-to-one communications in wireless sensor networks. Computer Communications, 30(4):863–872, February 2007.
[32] Aristides Mpitziopoulos, Damianos Gavalas, Charalampos Konstantopoulos, and Grammati Pantziou. Mobile Agent Middleware for Autonomic Data Fusion in Wireless Sensor Networks, chapter Autonomic Computing and Networking, pages 57–81. Springer, 2009.
[33] Hyacinth S. Nwana. Software agents: An overview. Knowledge Engineering Review, 11(3)(205-44), October/November 1996.
[34] Hairong Qi, S. Sitharama Iyengar, and Krishnendu Chakrabarty. Multiresolution data integration using mobile agents in distributed sensor networks. IEEE Transactions on Systems, Man, and Cybernetics, 31(3):383–91, 2001.
[35] Hairong Qi and Feiyi Wang. Optimal itinerary analysis for mobile agents in ad hoc wireless sensor networks. In the proceedings of the13th International Conference on Wireless Communication, Calgary, Alberta, Canada, July 9-11, pages 147–153, 2001.
[36] Hairong Qi, Yingyue Xu, and Xiaoling Wang. Mobile-agent-based collaborative signal and information processing in sensor networks. Proceedings of the IEEE, 91(8):1172–83, 2003.
[37] Ramesh Rajagopalan and Pramod K. Varshney. Data aggregation techniques in sensor networks: A survey. IEEE Communications Surveys and Tutorials, 8:48–63, 2006.
[38] Marcelo G. Rubinstein, Igor M. Moraes, Miguel Elias M. Campista, Lu Henrique M. K. Costa, and Otto Carlos M. B. Duarte. A survey on wireless ad hoc networks. Mobile and Wireless Communication Networks, Santiago, Chile, August 20-25, pages 1–33, 2006.
[39] Yu-Chee Seng, Sheng-Po Kuo, Hung-Wei Lee, and Chi-Fu Huang. Location tracking in a wireless sensor network by mobile agents and its data fusion strategies. 2nd international conference on Information processing in sensor networks (IPSN'03), Palo Alto, California, USA, April 22-23, pages 625–641, 2003.
[40] PREETI SETHI, DIMPLE JUNEJA, and NARESH CHAUHAN. A mobile agent-based event driven route discovery protocol in wireless sensor network: Aerdp. International Journal of Engineering Science and Technology (IJEST), 3(12):8422–9, December 2011.

[41] Mohamed A. Sharaf, Jonathan Beaver, Alexandros Labrinidis, and Panos K. Chrysanthis. Tina: a scheme for temporal coherency-aware in-network aggregation. Third ACM International Workshop on Data Engineering for Wireless and Mobile Access, MobiDE 2003, San Diego, California, USA, September 19, pages 69–76, 2003.
[42] Kazem Sohraby, Daniel Minoli, and Taien Znati. Wireless Sensor Network Technology, Protocols and Applications. John Wiley & Sons, Inc., 2007.
[43] Xue Wang, Dao wei Bi, Liang Ding, and Sheng Wang. Agent collaborative target localization and classification in wireless sensor networks. Sensors, 7:1359–1386, 2007.
[44] Johanna Williams. Telosb sensor network helpful in many different applications, 2014. http://telosbsensors. wordpress.com/, Retrieved (April 2014).
[45] Qishi Wu, Nageswara S. V. Rao, Jacob Barhen, S. Sitharama Iyengar, Vijay K. Vaishnavi, Hairong Qi, and Krishnendu Chakrabarty. On computing mobile agent routes for data fusion in distributed sensor networks. IEEE TRANSACTIONS ON KNOWLEDGE AND DATA ENGINEERING, 16:740–753, 2004.
[46] Jiuqiang Xu, Wei Liu, Fenggao Lang, Yuanyuan Zhang, and Chenglong Wang. Distance measurement model based on RSSI in WSN. Wireless Sensor Network, 2(8):606–611, 2010.
[47] Yingyue Xu and Hairong Qi. Mobile agent migration modeling and design for target tracking in wireless sensor networks. Ad Hoc Networks, 6(1):1–16, 2008.
[48] Zhe Yang and Abbas Mohammed. A survey of routing protocols of wireless sensor networks. Technical report, Blekinge Institute of Technology Sweden, 2010.
[49] Xiangbin Zhu and Wenjuan Zhang. A mobile agent-based clustering data fusion algorithm in WSN. International Journal of Electrical and Computer Engineering, 5(5):227–280, 2010.

## Authors

**Saeid Pourroostaei Ardakani** received his MSc. degree in computer engineering from Iran University Science and Technology (IUST) at 2007. In 2010, he joined university of bath, UK to do a PhD in computer science. Saeid obtained his PhD at 2014 focusing on Wireless Sensor Network intelligent routing protocols. Currently, he works in Allameh Tabataba'i University (Tehran, Iran) as an assistant professor in Computer Science. Saeid's research is not limited to but mainly focuses on WSNs, Internet of Things applications (IOT), Ubiquitous Computing and Cloud Computing.

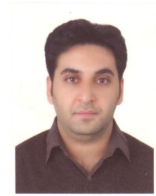